\documentclass{osa-article}

\journal{osac}

\articletype{Research Article}

\usepackage{lineno}

\begin{document}

\title{Controlling periodic Fano resonances of quantum acoustic waves with a giant atom coupled to microwave waveguide}

\author{Po-Chen Kuo,\authormark{1,2} Jhen-Dong Lin,\authormark{1,2} Yin-Chun Huang,\authormark{1} and Yueh-Nan Chen\authormark{1,2,*}}

\address{\authormark{1}Department of Physics, National Cheng Kung University, 701 Tainan, Taiwan\\
\authormark{2}Center for Quantum Frontiers of Research and Technology, NCKU, 701 Tainan, Taiwan}

\email{\authormark{*}yuehnan@mail.ncku.edu.tw} 



\begin{abstract}
Nanoscale Fano resonances, with applications from telecommunications to ultra-sensitive biosensing, have prompted extensive research. We demonstrate that a superconducting qubit, jointly coupled to microwave waveguides and an inter-digital transducer composite device, can exhibit acoustic Fano resonances. Our analytical framework, leveraging the Taylor series approximation, elucidates the origins of these quantum acoustic resonances with periodic Fano-like interference. By analyzing the analytical Fano parameter, we demonstrate that the Fano resonances and their corresponding Fano widths near the resonance frequency of a giant atom can be precisely controlled and manipulated by adjusting the time delay. Moreover, not just the near-resonant Fano profiles, but the entire periodic Fano resonance features can be precisely modulated from Lorentz, Fano to quasi-Lorentz shapes by tuning the coupling strength of the microwave waveguide. Our analytical framework offers insights into the control and manipulation of periodic Fano resonances in quantum acoustic waves, thereby presenting significant potential for applications such as quantum information processing, sensing, and communication.
\end{abstract}

\section{Introduction}
Fano resonance is a phenomenon that occurs when a discrete quantum state interferes with a continuum of states, resulting in an asymmetric spectral line shape~\cite{Yuri2010}. Fano resonance has been observed in various physical systems, such as atomic and molecular physics~\cite{Lambert2006,Paliwal2021}, plasmonics~\cite{Lukyanchuk2010,GuangYin2011,Shafiei2013,Kuo2016}, metamaterials~\cite{Lukyanchuk2010}, and photonics~\cite{Limonov2017}. Investigating Fano resonance, especially in quantum systems, is important for several reasons. First, Fano resonance reveals the underlying quantum interference mechanism that governs the interaction between discrete and continuous states~\cite{Limonov2021}. Second, Fano resonance can be used to manipulate light-matter interactions at the nanoscale, and results in enhancing optical absorption, emission~\cite{Zhen2013,Krger2018}, and scattering~\cite{Kuo2016}. Third, Fano resonance enables novel applications in sensing~\cite{Romain2019,GuangYin2017}, lasing~\cite{Zhen2013}, switching~\cite{Stern2014,Dabidian2015,Wang2009,JhenDong2022}, and nonlinear and slow-light devices~\cite{Wenyuan2020,Gennady2011}. Therefore, Fano resonance is a fundamental and versatile concept that has implications for both basic and applied physics~\cite{Yuri2010,Lukyanchuk2010,Limonov2017}.

Surface acoustic waves (SAWs), generated through piezoelectric acousto-electric transducers like interdigital transducers (IDTs) and propagating along elastic materials such as piezoelectric crystals~\cite{Manenti2017}, can be used to create a one-dimensional waveguide for phonons. This waveguide can be coupled to a superconducting qubit, allowing for the controlled emission and absorption of phonons~\cite{Forsch2019}. Compared to microwave photons, waveguide quantum electrodynamics (QED) with SAWs present several advantageous features. The significantly reduced propagation speeds of these waves, which are approximately five orders of magnitude slower than light, provide an extended interaction time frame, prolonged decoherence periods, and minimal losses within the GHz frequency range when interacting with quantum systems~\cite{Andersson2019,Bienfait2019}. These characteristics offer enriched opportunities for precise manipulation and control of quantum states~\cite{Satzinger2018}. Additionally, the shorter SAW-phonon wavelength at equivalent frequencies to microwave photons, resulting in smaller acoustic mode volumes, allows for enhanced acoustic coupling to superconducting qubits compared to electrical coupling~\cite{Chu2017,Delsing2019}. Collectively, these features position SAWs as a promising alternative to microwave photons, with significant potential for quantum sensing~\cite{McKibben2023}, quantum information processing~\cite{Byeon2021}, communication~\cite{Satzinger2018}, and overall advancements in quantum devices~\cite{FriskKockum2022}.

The integration of Fano resonance with SAWs has garnered considerable interest in recent years, owing to its prospective applications across a multitude of fields including, but not limited to, coupled resonators~\cite{Hideo2020}, topological systems~\cite{Romain2019}, quantum devices~\cite{kitzman2023}, metasurfaces~\cite{Jin2017,Jian2021}, and phonon laser~\cite{Tang2009}. This synergy between Fano resonance and SAWs paves the way for novel functionalities and unattainable performance enhancements when either phenomenon operates in isolation. Such advancements include improving the sensitivity~\cite{Hideo2020} and robustness~\cite{Romain2019} of SAW-based sensors, overcoming the narrow bandwidth constraints of Fano resonance~\cite{kitzman2023}, managing potential interactions between hypersound and charge carriers~\cite{Vasileiadis2020}, and manipulating the transmission and reflection coefficients of acoustic waves~\cite{Jin2017,Zaki2020}.

However, the existing body of work primarily concentrates on the demonstration of a single Fano resonance, leaving the study of periodic Fano resonance features in SAWs comparatively unexplored. The ability to control the periodic Fano resonance of quantum acoustics is an area of research that has not been extensively investigated. Within this context, we propose a system incorporating a "giant atom" jointly coupled to a microwave and an acoustic wave waveguide. This configuration unveils the periodic Fano resonance features in the surface acoustic wave (SAW) scattering spectra and suggests a methodology for controlling and fine-tuning these resonances. In section~\ref{sec2}, we first introduce our model Hamiltonian and lay out the theoretical foundation for determining the scattering spectra for both microwave and SAW. Subsequently, in section~\ref{sec3}, we present our analytical findings and discuss the behavior of periodic Fano resonances within the scattering spectra, employing Taylor series expansion. Moreover, leveraging the analytical Fano parameter, we delve into the impact of varying parameters such as intrinsic time delay and coupling strength on the periodic Fano resonances. This investigation underscores the ability to control the acoustic Fano characteristics within the scattering spectra.

\section{Models and methods}\label{sec2}
\subsection{Model Hamiltonian}
\begin{figure}[h!]
\centering\includegraphics[width=11cm]{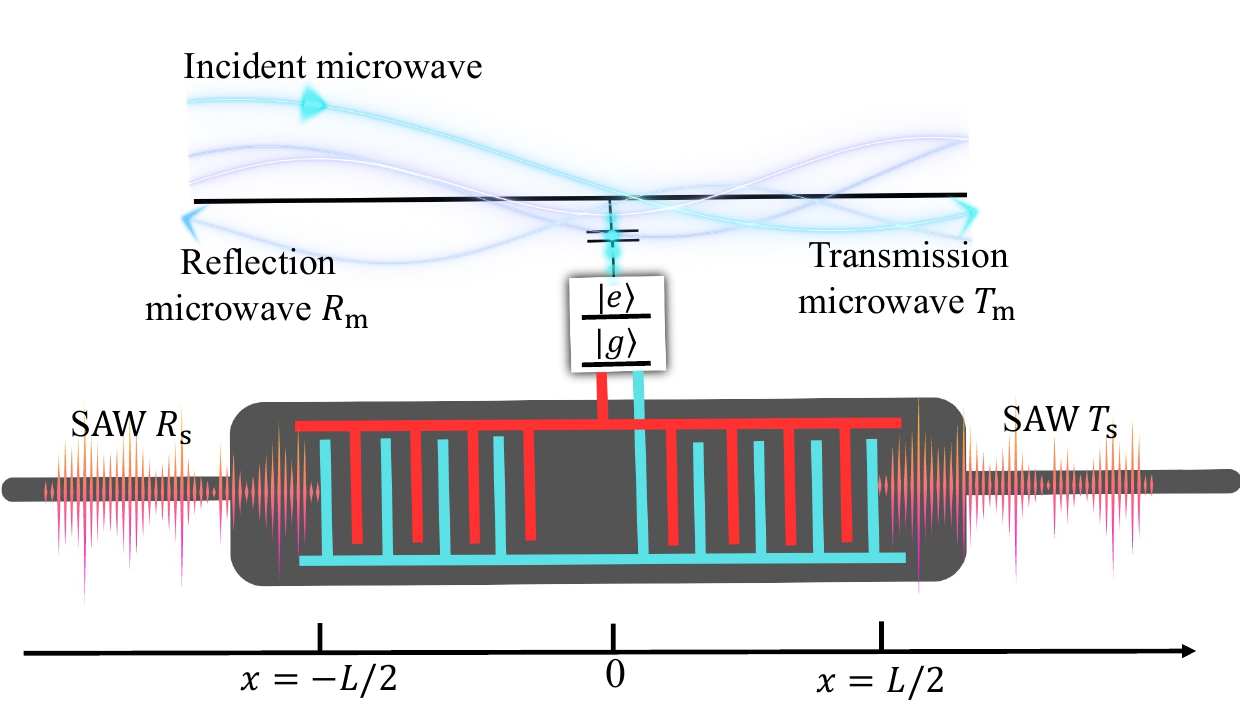}
\caption{The proposed system designed to produce periodic Fano resonance in quantum SAW. It comprises a microwave waveguide (upper black transmission line) interfaced with a superconducting qubit. The latter is modeled as a two-level system with states $|e\rangle$ and $|g\rangle$, effectively serving as a giant atom. This giant atom setup is achieved by bifurcating the IDT (red and green interdigitated finger structure) into two electrically linked coupling points (with the left point positioned at $x=-L/2$ and the right one at $x=L/2$), separated by distance $L$. Interaction between the incident microwave photon and the superconducting qubit induces transitions in the giant atom, causing SAW phonon scattering through the IDT and propagating in both left ($R_{\text{s}}$) and right ($T_{\text{s}}$) directions along the piezoelectric substrate. Concurrently, the incident microwaves are reflected $(R_{\text{m}})$ and transmitted $(T_{\text{m}})$.}\label{fig0}
\end{figure}

Here, we consider a superconducting qubit, which functions as a "giant" two-level atom (a). This qubit is simultaneously connected to a superconducting transmission line for microwave photon conveyance and the IDTs for propagating SAW generated on a piezoelectric substrate through the piezoelectric effect, as depicted in Fig.~\ref{fig0}. The total (T) Hamiltonian in our model, therefore, can be characterized as
\begin{equation}
\begin{aligned}
H_{\text{T}} = H_{\text{a}}+H_{\text{m}}+H_{\text{s}}+H_{\text{am}}+H_{\text{as}}.
\end{aligned}
\end{equation}
The Hamiltonian, which describes the giant two-level atom with the bare frequency ($\omega_{e}\approx 2.3$ (GHz)~\cite{Andersson2019}) of the superconducting qubit's excited state, is denoted as
\begin{equation}
\begin{aligned}
H_{\text{a}} = \hbar\omega_{e}\sigma_{ge}^{\dagger}\sigma_{ge},
\end{aligned}
\end{equation}
where $\sigma_{ge}=|g\rangle\langle e|$ ($\sigma_{ge}^{\dagger}=|e\rangle\langle g|$) with $|e\rangle$ ($|g\rangle$) being the excited (ground) state. The propagation of microwave (m) photons on the superconducting transmission line can be described by the model Hamiltonian~\cite{Chang2007,GuanTing2018,ChiaYi2020}
\begin{equation}
\begin{aligned}
H_{\text{m}} = \hbar\sum_{\alpha=\text{r},\text{l}}\int\omega_{\text{m}}a^{\dagger}_{\alpha\omega_{\text{m}}}a_{\alpha\omega_{\text{m}}}d\omega_{\text{m}},
\end{aligned}
\end{equation}
where $a^{\dagger}_{\alpha\omega{\text{m}}} (a_{\alpha\omega_{\text{m}}})$ represents the creation (annihilation) of right-($\alpha=\text{r}$) or left-($\alpha=\text{l}$) propagating microwave photons, with frequency $\omega_{\text{m}}$. The interaction between the superconducting qubit and the propagating microwave photon, with the coupling strength $g_{\text{m}}$, is given by~\cite{Fan2005,Chang2007,GuangYin2014,Lambert2016}:
\begin{equation}
\begin{aligned}
H_{\text{am}} = \hbar\sum_{\alpha=\text{r},\text{l}}\int g_{\text{m}}\Big[a^{\dagger}_{\alpha\omega_{\text{m}}}\sigma_{ge}+\text{H.c.}\Big]d\omega_{\text{m}},
\end{aligned}
\end{equation}
where H.c. denotes the Hermitian conjugate. Meanwhile, the propagating SAW (s), described by the piezoelectric transmission line Hamiltonian
\begin{equation}
\begin{aligned}
H_{\text{s}} = \hbar\sum_{\alpha=\text{r},\text{l}}\int \omega_{\text{s}}b^{\dagger}_{\alpha\omega_{\text{s}}}b_{\alpha\omega_{\text{s}}}d\omega_{\text{s}},
\end{aligned}
\end{equation}
is capacitively coupled to the superconducting qubit through the interaction Hamiltonian~\cite{Johansson2017,Andersson2019}
\begin{equation}
\begin{aligned}
H_{\text{as}} = \hbar\sum_{\alpha=\text{r},\text{l}}\int g_{\text{s}}\Big[\phi_{\text{s}}b^{\dagger}_{\alpha\omega_{\text{s}}}\sigma_{ge}+\text{H.c.}\Big]d\omega_{\text{m}},
\end{aligned}
\end{equation}
where $b^{\dagger}_{\alpha\omega{\text{s}}} (b_{\alpha\omega_{\text{s}}})$ piezoelectrically creates (annihilate) right-($\alpha=\text{r}$) or left-($\alpha=\text{l}$) propagating SAW with frequency $\omega_{\text{s}}$. Since SAWs propagate slower than microwave photons, their wavelength $(\lambda_{s})$ is relatively small, comparable to, or even a few times smaller than the distance (L) between the two coupling points on the IDTs. This smaller wavelength introduces a significant intrinsic time delay, which is defined as $\phi_{\text{s}}=e^{i k_{s} L/2}+e^{-i k_{s} L/2}$. In this equation, the wave vector of the SAW ($k_{s}$) is given by $\omega_{\text{s}}/v_{\text{g}}=2\pi/\lambda_{s}$, where $v_{\text{g}}$ is the traveling velocity of the SAW on the piezoelectric substrate. Because of this relationship, the superconducting qubit can act as a “giant atom” during its interaction with SAW.

\subsection{Scattering spectrum within the framework of a single-excitation process}
In this investigation, we delve into the complex dynamics of a giant atom specifically in the weak coupling regime, where $\gamma_{\text{m}},\gamma_{\text{s}}\ll\omega_{e}$. Within this regime, we can make use of the rotating wave approximation and the single-excitation approximation. These approximations allow us to simplify the system and make it analytically tractable. Thus, the total state of our quantum system, composed of the giant atom, microwave, and the SAW field, can be represented as~\cite{Fan2005,Chang2007,Johansson2017}
\begin{equation}
\begin{aligned}
|\psi(t)\rangle=\sum_{\alpha=\text{l},\text{r}}&\Big[
\int d\omega_{\text{m}}c_{\alpha\omega_{\text{m}}}(t)a^{\dagger}_{\alpha\omega_{\text{m}}}
+\int d\omega_{\text{s}}c_{\alpha\omega_{\text{s}}}(t)b^{\dagger}_{\alpha\omega_{\text{s}}}
\Big]|g,0_{\alpha\omega_{\text{m}}},0_{\alpha\omega_{\text{s}}}\rangle\\
&+e(t)|e,0_{\alpha\omega_{\text{m}}},0_{\alpha\omega_{\text{s}}}\rangle,
\end{aligned}
\end{equation}
where $c_{\alpha\omega_{\text{m}}}(t)$, $c_{\alpha\omega_{\text{s}}}(t)$, and $e(t)$ represent the probability amplitudes of the propagating microwave photon, SAW phonon, and the excited state, respectively. For the left-orientated ($\alpha=\text{l}$) states, these represent the Fock state of microwave photon $|g,1_{\text{l}\omega_{\text{m}}},0_{\alpha\omega_{\text{s}}}\rangle$ and the SAW phonon $|g,0_{\alpha\omega_{\text{m}}},1_{\text{l}\omega_{\text{s}}}\rangle$. Similarly, for the right-orientated ($\alpha=\text{r}$) states, they represent the Fock state of microwave photon $|g,1_{\text{r}\omega_{\text{m}}},0_{\alpha\omega_{\text{s}}}\rangle$ and the SAW phonon $|g,0_{\alpha\omega_{\text{m}}},1_{\text{l}\omega_{\text{s}}}\rangle$. By solving the Schr\"{o}dinger equation $H_{\text{T}}|\psi(t)\rangle=i\hbar\partial_{t}|\psi(t)\rangle$, one can obtain the system's equations of motion~\cite{Johansson2017}. We first examine the long-time behavior of the quantum system, i.e., $t \to \infty$. This approach is particularly advantageous for experimental accessibility, especially when detecting the reflected and transmitted microwave photons as well as SAW phonons. The initial condition presumes the injection of a singular microwave photon into the transmission line, such that $c_{\alpha\omega_{\text{m}}}(0)=1$ and $c_{\alpha\omega_{\text{s}}}(0)=0$. Consequently, the IDT generates a SAW. Therefore, we are able to ascertain the transmission ($T$) and reflection ($R$) spectra for both microwave photon (m) and SAW phonon (s) in the long-time limit
\begin{equation}
\begin{aligned}
R_{\text{m}}=\lim_{t \to \infty}|c_{\text{l}\omega_{\text{m}}}(t)|^{2}
=\frac{\gamma _{\text{m}}^2}{\left| \gamma _{\text{m}}+\left(1+e^{i \omega_{\text{m}} t_{L}  }\right)
   \gamma _{\text{s}}-i \left(\omega_{\text{m}} -\omega _e\right)\right| {}^2},\label{Rm}
\end{aligned}
\end{equation}

\begin{equation}
\begin{aligned}
R_{\text{s}}=T_{\text{s}}=\lim_{t \to \infty}|c_{\alpha\omega_{\text{s}}}(t)|^{2}
=\frac{\gamma _{\text{m}} \gamma _{\text{s}} \left[\cos (\omega_{\text{s}} t_{L})+1\right]}{\left| \gamma _{\text{m}}+\left(1+e^{i \omega_{\text{s}} t_{L}}\right)
   \gamma _\text{s}-i \left(\omega_{\text{s}} -\omega _{\text{e}}\right)\right| {}^2},\label{Rs}
\end{aligned}
\end{equation}

\begin{equation}
\begin{aligned}
T_{\text{m}}=1-R_{\text{m}}-2R_{\text{s}}.
\end{aligned}
\end{equation}

\section{Results and discussion}\label{sec3}
\subsection{Analytical analysis for the periodic Fano resonances}
\begin{figure}[h!]
\centering\includegraphics[width=12cm]{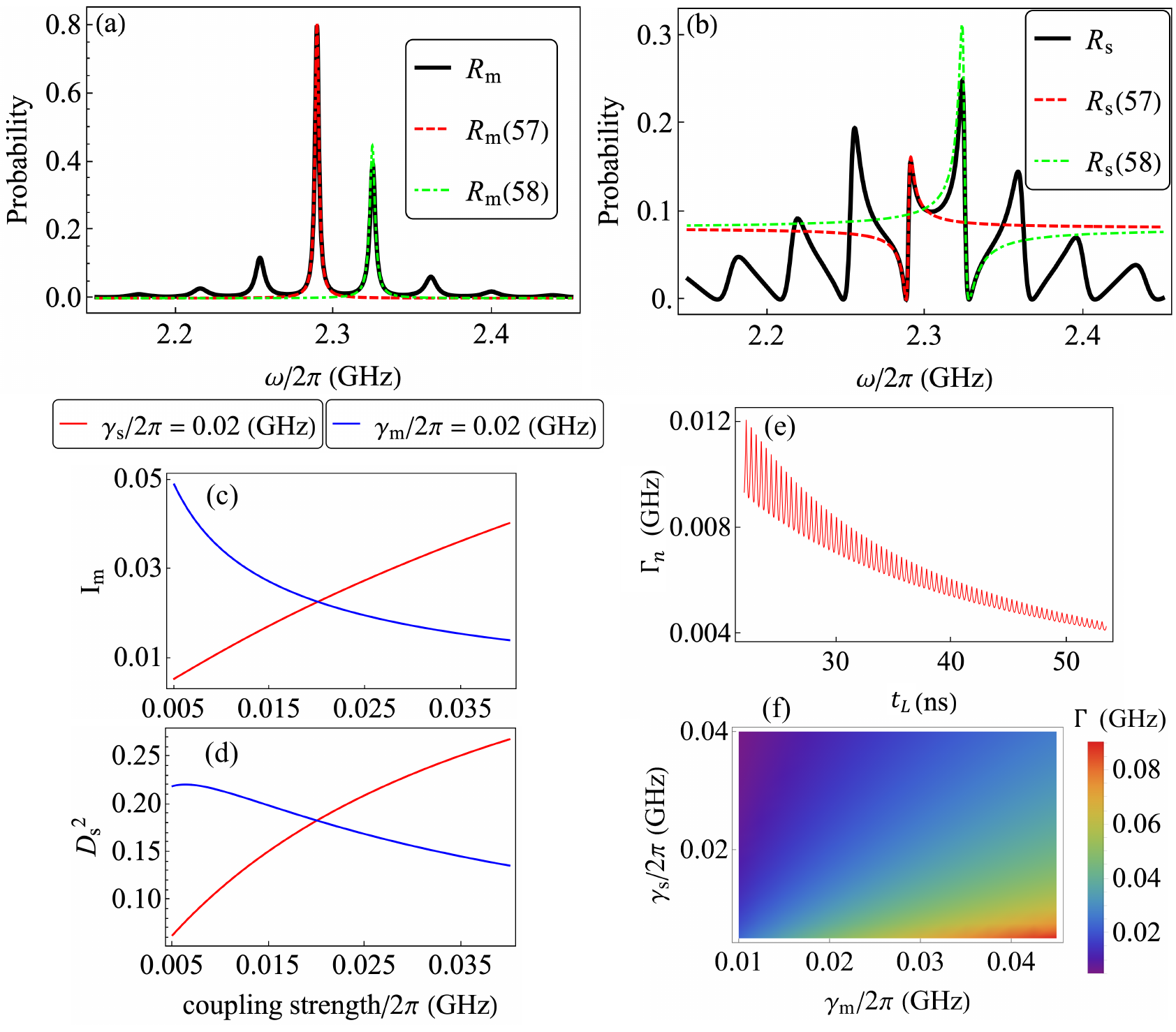}
\caption{The scattering (reflectance) spectra of microwave and SAW. (a) The reflection probability $R$ (black-solid curve) of (a) microwave (m) and (b) SAW (s) as functions of microwave and SAW field frequency ($\omega$). The red and green dashed curves display the Taylor approximation of the reflection probabilities at $n=57$ and $n=58$, closely aligning with the periodic Lorentzian-like lineshape $R_{\text{m}}$ and Fano-like lineshape $R_{\text{s}}$. The red dashed curve, with a resonant frequency near $\omega_{e}$, exhibits a closer alignment with the exact reflection probability. An overview of the intensities, with (c) representing $I_{\text{m}}$ and (d) representing $D_{\text{s}^{2}}$, are plotted against the coupling strengths $\gamma_{\text{m}}$ (red solid line) and $\gamma_{\text{s}}$ (blue solid line). (e) The resonance width ($\Gamma_{n}$) variation with respect to the time delay $t_{L}$. (f) The density plot of $\Gamma_{n}$ against $\gamma_{\text{m}}$ and  $\gamma_{\text{s}}$}\label{fig1}
\end{figure}

When we initially introduce the microwave photon into the microwave waveguide, the scattering spectrum, specifically $R_{\text{m}}$, discloses the periodic peaks that follow a Lorentzian distribution as illustrated in Fig.~\ref{fig1}~(a). To analytically dissect the emergence of these periodic Lorentzian peaks, we resort to a Taylor or Pad\'{e} approximation~\cite{Jie2010,Mauro2023} of the exponential function or trigonometric function. In this context, a second-order Taylor expansion is more fitting to delineate this characteristic periodicity. Hence, we have $e^{i\omega t_{L}}\approx\frac{1}{2} t_{L}^2 (\omega -\omega_{n}){}^2-i t_{L} (\omega-\omega_{n})-1$, where $\omega_{n}=2\pi(n+1)/t_{L}$ denotes the $n$-th periodic frequency that triggers the destructive interference of the SAW with the regulated time delay $t_{L}=L/v_{\text{gs}}$. Here, $v_{\text{gs}}$ stands for the group velocity of the SAW. Incorporating this Taylor expansion into Eq.(\ref{Rm}), we can derive the analytical Lorentzian form of $R_{\text{m}}(n)$ at the $n$-th periodic frequency, which aligns well with the Lorentzian peaks of $R_{\text{m}}$ as shown in Fig.\ref{fig1}(a). It's given by the following equation
\begin{equation}
\begin{aligned}
R_{\text{m}}(n)=I_{\text{m}} \frac{\Gamma_{n}}{\left(\omega -\omega _{\text{eff},n}\right){}^2+\Gamma_{n}^2},
\end{aligned}
\end{equation}
where $I_{\text{m}}$, $\Gamma_{n}$, and $\omega_{\text{eff},n}$ represent the intensity of the $n$-th Lorentz-type peak, $n$-th resonance width, and effective frequency of the $n$-th resonance peak, respectively. It should be noted that the width of the peak, denoted as $\Gamma_{n}$ [also known as the full width at half maximum (FWHM)], can be manipulated by adjusting both the intrinsic time delay $t_{L}$ and the coupling strengths $(\gamma_{\text{m}}$,$\gamma_{\text{s}})$. This width is determined by the parameters $\eta$ and $\eta_{n}$
\begin{equation}
\begin{aligned}
\Gamma_{n} = \frac{\eta _n}{\eta },\label{Gamma}
\end{aligned}
\end{equation}
where $\eta = 2 \gamma _{\text{m}} \gamma _{\text{s}} t_{L}^2 +2 \left(\gamma _{\text{s}}t_{L}+1\right){}^2$ and $\eta_{n} = \sqrt{2\gamma _{\text{m}}\left[2 \gamma _{\text{s}}t_{L}^2 \left(\omega_e-\omega_{n}\right){}^2+\gamma_{\text{m}}\eta \right]}$. The intensity of the reflection spectrum of the microwave photon is characterized by $I_{\text{m}}$, given by:
\begin{equation}
\begin{aligned}
I_{\text{m}} = \frac{2 \gamma _{\text{m}}^2}{\eta_n},
\end{aligned}
\end{equation}
where increasing $\gamma _{\text{m}}$ enhances the strength of the spectrum. However, increasing $\gamma _{\text{s}}$ has the opposite effect, reducing the intensity. This reduction occurs because an increase in $\gamma _{\text{s}}$ results in the partial transfer of energy from the microwave photon to the SAW phonon through the IDT, as depicted in Fig.~\ref{fig1}(c). The effective resonance frequency $\omega_{\text{eff},n}$, encoded as $\omega_{n}$, exhibits periodicity, reflecting the periodic nature of the resonance in the spectrum
\begin{equation}
\begin{aligned}
\omega_{\text{eff},n} = \omega_{n} +\frac{2}{\eta}\left(\omega_e-\omega_{n}\right) \left(\gamma _{\text{s}}t_{L}+1\right).
\end{aligned}
\end{equation}
Crucially, upon scrutinizing the scattering spectra of the SAW, generated via the IDT coupling to the giant atom, we discern a fascinating feature, specifically, the manifestation of periodic Fano resonances as depicted in Fig.~\ref{fig1}(b). Proceeding with the same methodology to decompose $R_{\text{s}}$ via a second-order Taylor expansion, the $R_{\text{s}}$ associated with the $n$-th Fano profile can be expressed as 
\begin{equation}
\begin{aligned}
R_{\text{s}}(n)=D_{\text{s}}^2 \frac{\left(\Omega _{\text{s}}+q\right){}^2}{\Omega _{\text{s}}^2+1}.
\end{aligned}
\end{equation}
In this equation, $\Omega _{\text{s}}$ and $q$ symbolize the reduced frequency normalized by the resonance width $\Gamma_{n}$
\begin{equation}
\begin{aligned}
\Omega _{\text{s}}=\frac{\omega -\omega _{\text{eff},n}}{\Gamma_{n} },
\end{aligned}
\end{equation}
and the Fano parameter, respectively
\begin{equation}
\begin{aligned}
q=\frac{2}{\eta _n} \left(\omega_e-\omega_{n}\right) \left(\gamma_{\text{s}}t_{L}+1\right).
\end{aligned}
\end{equation}
The latter captures the degree of asymmetry present in the Fano resonance. 

The overall intensity of the spectra, encompassing the Fano profiles, can be expressed as
\begin{equation}
\begin{aligned}
D_{\text{s}}^{2}=\frac{\gamma _{\text{m}} \gamma _{\text{s}} t_{L}^2 }{\eta},
\end{aligned}
\end{equation}

This can be amplified by increasing $\gamma_{\text{m}}$. However, the increase in $\gamma_{\text{s}}$ gradually reduces the intensity due to the even redistribution of the intensity across each periodic Fano profile, as illustrated in Fig.~\ref{fig1}~(d). It should be noted that the Fano resonance in $R_{\text{s}}$ can be more precisely characterized by $R_{\text{s}}(n)$ when the defined $n$-th Fano resonance approaches $\omega_{e}$. Simultaneously, the resonance width ($\Gamma_{n}$) can be modulated by variously designed $t_{L}$ and the alterations in coupling strengths.

As we extend $L$ and consequently enlarge $t_{L}$, the intensity of the scattering spectra experiences substantial alterations corresponding to the variations in the frequency of the SAW, in accordance with Eq.(\ref{Rs}). This leads to a decreased resonance width ($\Gamma_{n}$), as illustrated in Fig.\ref{fig1}(e). Interestingly, increasing $\gamma_{\text{m}}$ enlarges $\Gamma_{n}$. In contrast, an increment in $\gamma_{\text{s}}$ decreases $\Gamma_{n}$, since $\Gamma_{n}$ is effectively proportional to $\sqrt{\gamma_{\text{m}}}$, but inversely proportional to $\gamma_{\text{s}}$, according to Eq.(\ref{Gamma}), as shown in Fig.\ref{fig1}(f).

\begin{figure}[h!]
\centering\includegraphics[width=9cm]{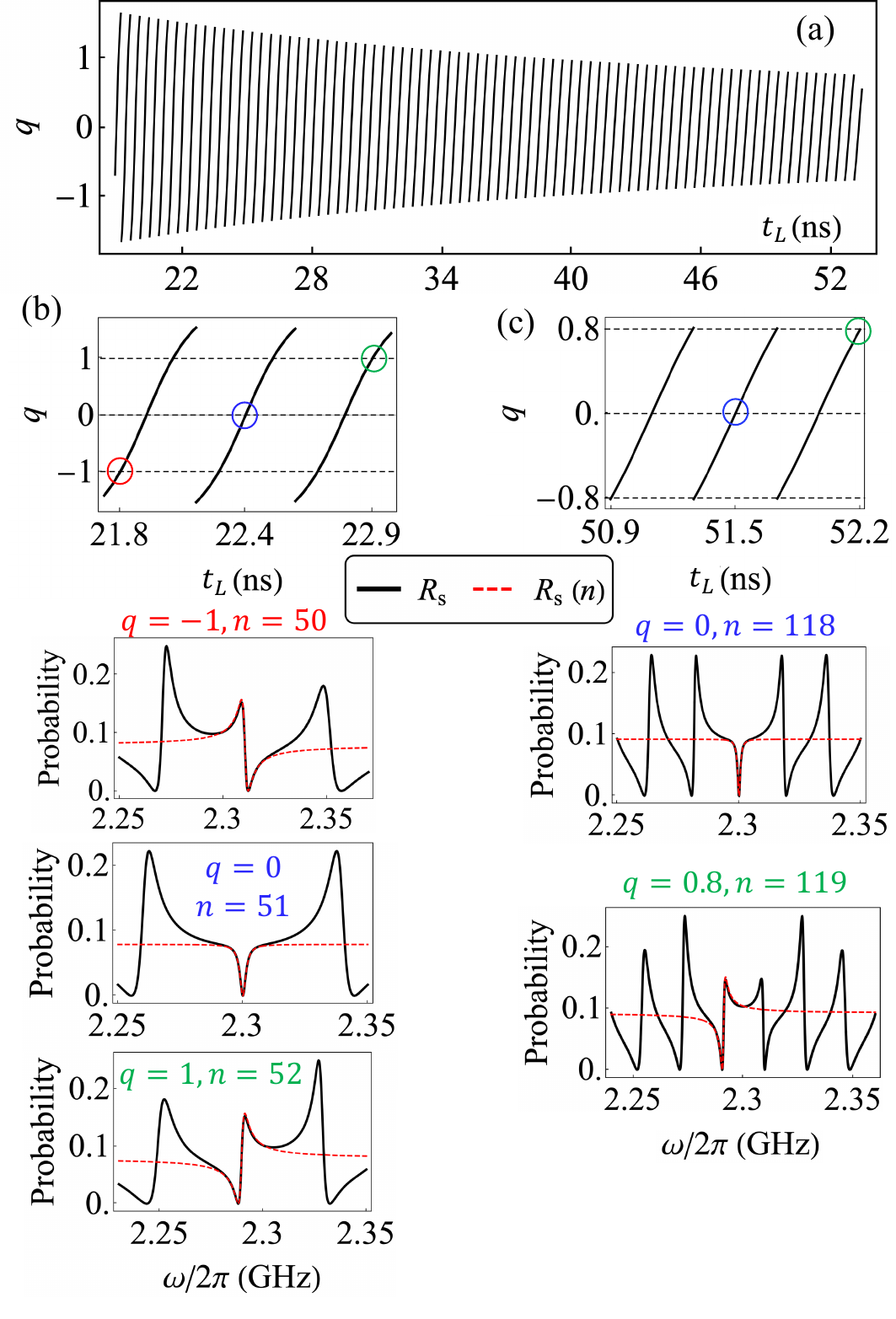}
\caption{(a) The variations of the Fano parameter $q$ with time delay $t_{L}$. (b) Extracting the shorter $t_{L}$ regime for periodic $q$ where $R_{\text{s}}$ transitions through Fano, quasi-Lorentz, and back to Fano profiles with respect to $t_{L}$ resulting in $q=-1,n=50$ (red circles), $q=0,n=51$ (blue circles), and $q=1,n=52$ (green circles). (c) Extracting the longer $t_{L}$ regime for $q$ with $R_{\text{s}}$ reveals sharper quasi-Lorentz and Fano profiles when $t_{L}$ is respectively tuned to yield $q=0,n=118$ (blue circles) and $q=0.8,n=119$ (green circles). }\label{fig2}
\end{figure}

\subsection{Tunable control of the periodic Fano resonances}
The Fano parameter $q$, also known as the Fano asymmetry factor, is a dimensionless quantity that characterizes the shape of a resonance profile in quantum systems. It arises from the interference between a discrete state and a continuum of states, and it can vary from zero (quasi-Lorentzian profile) to infinity (Lorentzian profile)~\cite{Deng2015,Nodar2023,Tang2009}. The ability to control the Fano parameter is important for various applications in fundamental quantum processes, quantum sensing, and improved spectroscopic measurements with enhanced sensitivity as well as selectivity, as it can affect the transmission, reflection, absorption, and emission properties of quantum systems~\cite{Limonov2017,Gollwitzer2021}.

 
\subsubsection{Effects of the time delay variations on Fano resonances properties}
In this part of our study, we aim to control the Fano resonance of the quantum SAW by adjusting the Fano parameter. The time delay $t_{L}$, which can be expressed as $L/v_{\text{gs}}$, plays a crucial role in inducing the interference effect. This interference arises from the possibility that a phonon emitted into the acoustic waveguide at one coupling point of the qubit, due to spontaneous decay, may later be reabsorbed by the qubit at the second coupling point, which is separated by a distance $L$ from the first coupling point. With this understanding, we concentrate on adjusting the Fano profile of the SAW by varying $t_{L}$. By manipulating  $t_{L}$, we effectively modify the Fano parameter $q$, thus allowing for control over the Fano resonance characteristics exhibited by the SAW. By setting parameters to realistic values such as $\gamma_{\text{m}}/2\pi=0.01\text{GHz}, \gamma_{\text{s}}/2\pi=0.038\text{GHz}$, we observe that as $t_{L}$ ranges from small to large values, the Fano parameter $q$ demonstrates periodic fluctuations with the $\omega_{n}$ close to $\omega_{e}$, as illustrated in Fig.~\ref{fig2}~(a).

This periodicity of $q$ in relation to certain parameters is also shown across a wide range of systems where $q=\cot\delta$, with $\delta$ symbolizing the phase shift~\cite{Yuri2010}. Interestingly, in our system, we note that the Fano parameter $q$ displays larger amplitude periodic variations for smaller $t_{L}$, and the amplitude of these variations decreases as $t_{L}$ increases. We specifically focus on the case with smaller $t_{L}$. As depicted in Fig.~\ref{fig2}(b), the Fano parameter $q$ experiences periodic variations roughly between -1 and 1. Therefore, by adjusting $t_{L}$, we can manipulate the value of $q$. For example, we can set $t_{L}$ to achieve $q=1$ or $q=-1$, which follows the equation with $\Delta_{\text{ep}}=\omega_{e}-\omega_{n}$
\begin{equation}
\begin{aligned}
\Delta _{\text{ep}}^2 \left(\eta -4 t_{L}^2 \gamma _{\text{m}} \gamma_{\text{s}}\right)=\eta \gamma _{\text{m}}^2,\label{q1}
\end{aligned}
\end{equation}
where the most prominent Fano resonances in the scattering spectrum are observed. The difference between $q=1$ and $q=-1$ simply results in a mirror reflection of the Fano profile shape. Adjusting $t_{L}$ to achieve $q=0$ by fulfilling the condition,
\begin{equation}
\begin{aligned}
\Delta_{\text{ep}}=0,
\end{aligned}
\end{equation}
we clearly observe a dip in the scattering spectra near the giant atom's resonance frequency $\omega_{e}$. This dip is similar to the phenomenon described in other works~\cite{Limonov2017}, where the system's interaction with discrete energy can be significantly suppressed, thereby rendering this giant atom as an invisible discrete state in such instances.

As we increase $t_{L}$, we observe a decrease in the amplitude of the periodic variation of the Fano parameter $q$, ranging roughly from 0.8 to -0.8. Taking the derivative of $q$ with respect to $t_{L}$, or $\partial_{T}(q)=0$, one can ascertain the maximum absolute value of $q$, denoted as $q_{\max }=\max |q|$, by satisfying the following relationship:
\begin{equation}
\begin{aligned}
T \Delta_{\text{ep}} \gamma_{\text{m}} \gamma_{\text{s}}\left(\Delta_{\text{ep}}^2+\gamma_{\text{m}}^2\right)=0.\label{qmax}
\end{aligned}
\end{equation}
Despite $|q_{\max}|<1$ (approximately 0.8), distinct Fano resonances in the scattering spectrum are still observable as shown in Fig.~\ref{fig2}(c). Similarly, even when $q=0$, narrower dips within the spectrum are still discernible. These observations underscore the fact that we can indeed manipulate the characteristics of Fano resonances within the scattering spectra by adjusting the dimension of $L$ and the corresponding $t_{L}$.
 
\begin{figure}[h!]
\centering\includegraphics[width=9cm]{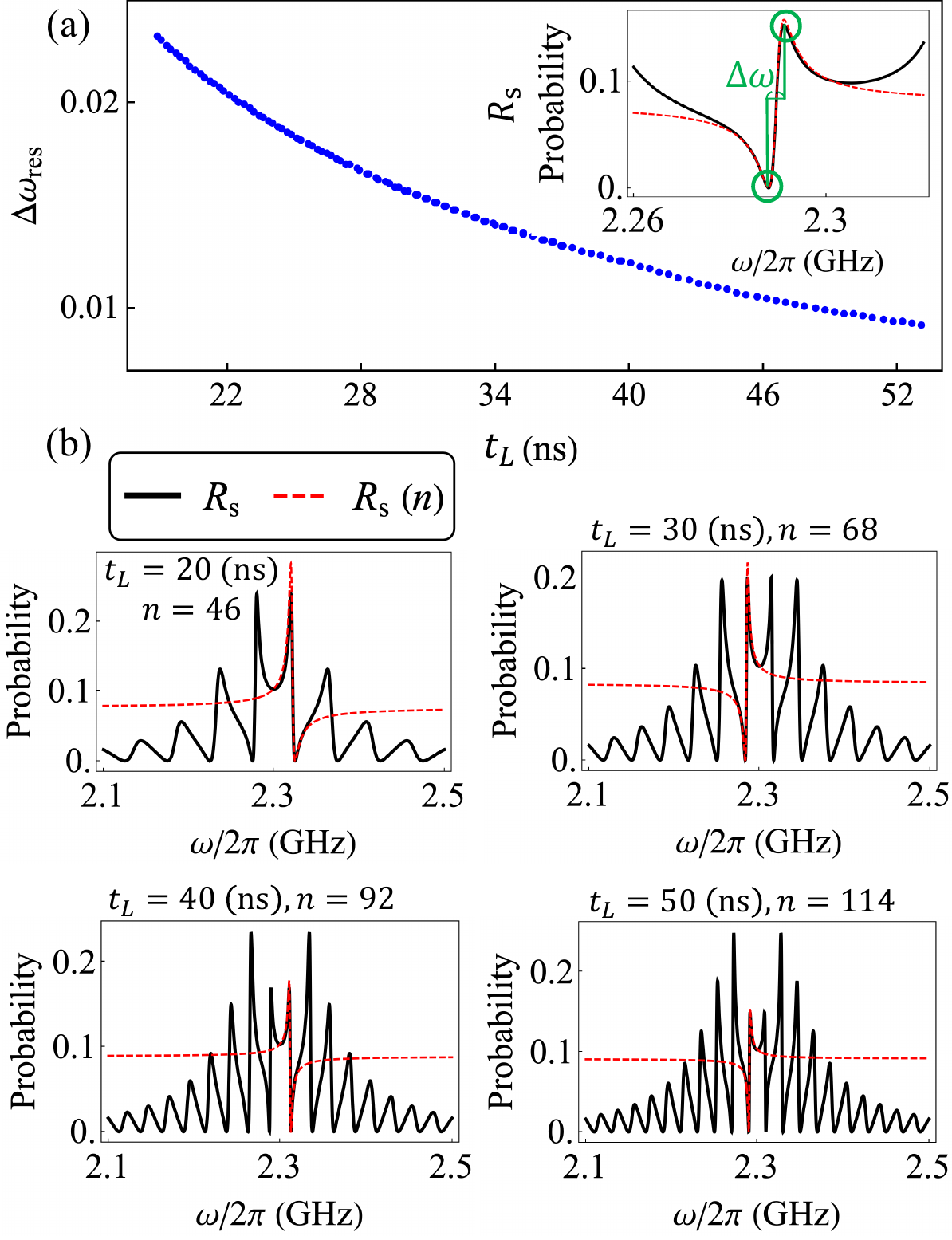}
\caption{(a) The variations in the width of the Fano resonances $\Delta\omega_{\text{res}}$, near the resonant frequency $\omega_{e}$, as a function of the time delay $t_{L}$. The inset illustrates the definition of the width of the Fano resonance. (b) A compression of Fano peaks (black-solid curve) is observed with an increase in time delay from $T=20$ (ns) to $T=50$ (ns), corresponding to a decrease in $\Delta\omega$. These near-resonant Fano peaks align well with $R_{\text{s}}(n)$ (red dashed) for n values ranging from 46 to 114.}\label{fig3}
\end{figure}

Besides the Fano parameter $q$, the width of the Fano resonance plays a crucial role in various applications, such as quantum switches and sensing. In the case of a quantum switch, a sharp Fano resonance with a narrow width is desirable as it allows for a significant change in transmission within a small range of wavelengths. This enables efficient operation of the switch with minimal energy consumption, as the state of the switch can be toggled by making small adjustments in the input wavelength~\cite{Yu2014,Stern2014,Dabidian2015,Rybin2015}. Similarly, a narrow width of the Fano resonance is preferred for sensing applications, as it not only enhances sensitivity but also improves the figure of merit~\cite{Lee2015}. Therefore, it is important to understand how to manipulate the width of the Fano resonance by tuning the system's parameters.

In addition to analyzing the variation of $q$ associated with Fano resonances, we can also manipulate the width of Fano resonances by adjusting $t_{L}$. Here, we characterize the Fano width ($\Delta\omega$) as the frequency disparity between the peak and the dip within the Fano resonance, as shown in the inset of Fig.~\ref{fig3}(a). The frequencies where the periodic dips occur can be denoted by $\omega_{n}$. The frequency corresponding to the periodic peak in the Fano profile, which approaches $\omega_{e}$, can be identified by setting $\partial_{\omega}[R_{\text{m}}(n)]=0$. Consequently, we can analytically derive the Fano width
\begin{equation}
\begin{aligned}
\Delta\omega = \frac{\Delta _{\text{ep}}^2+\gamma _{\text{m}}^2}{\Delta _{\text{ep}}
   \left(\gamma _{\text{s}}t_{L}+1\right)}.\label{dw}
\end{aligned}
\end{equation}
Our analysis of the results in Fig.\ref{fig3}(a) reveals a periodic behavior in $\Delta\omega$ with the variation of $t_{L}$. To accurately calculate $\Delta\omega$ of the Fano profile as a function of $t_{L}$, we adopt a two-step approach. If $q_{\max}>1$, we first identify specific values of $t_{L}$ that yield $|q|=1$, using Eq.(\ref{q1}). These values of $t_{L}$ are then incorporated into Eq.(\ref{dw}) to derive $\Delta\omega$. Alternatively, if $q_{\max}<1$, we directly utilize the $t_{L}$ value corresponding to $q_{\max}$ derived from Eq.(\ref{qmax}), and substitute it into Eq.~(\ref{dw}) to compute $\Delta\omega$.

As we increase $t_{L}$, we observe a gradual decrease in $\Delta\omega$, exhibiting a trend paralleling that shown in Fig.\ref{fig1}(e). Our findings suggest that increasing $t_{L}$ not only alters the Fano parameter but also leads to a more pronounced change in the spectra with respect to the SAW frequency ($\omega$). This results in the formation of densely packed periodic Fano profiles in the scattering spectra, corroborating the earlier observation made in Fig.\ref{fig2}(c). The progressive decrease in $\Delta\omega$ with the enhancement of $t_{L}$ fosters the manifestation of more concentrated and distinct Fano resonances, as depicted in Fig.~\ref{fig3}(b).

\begin{figure}[h!]
\centering\includegraphics[width=12cm]{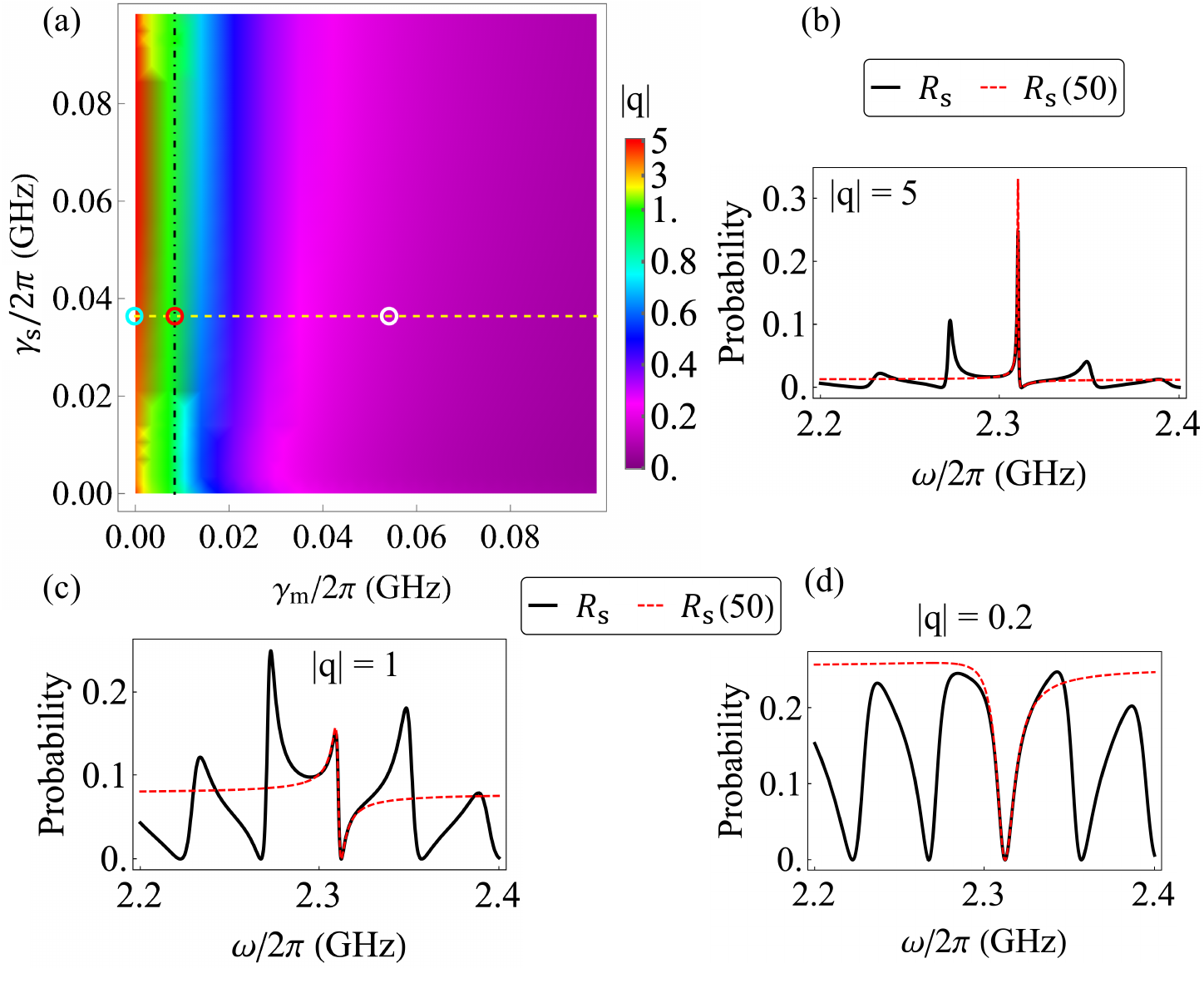}
\caption{(a) The density plot of the Fano parameter $q$ as a function of $\gamma_{\text{s}}$ and $\gamma_{\text{m}}$. For a fixed $\gamma_{\text{m}}/2\pi=0.01$ GHz, the Fano parameter remains relatively constant (black-dashed line for $|q|\approx 1$). When the microwave waveguide coupling strength is increased (with fixed $\gamma_{\text{s}}/2\pi=0.038$ GHz), different profiles of $R_{\text{s}}$ emerge: (b) Lorentz-like for $\gamma_{\text{m}}/2\pi\approx 0.001$ GHz (light blue circle), (c) Fano-like for $\gamma_{\text{m}}/2\pi= 0.01$ GHz (red circle), and (d) quasi-Lorentz for $\gamma_{\text{m}}/2\pi\approx 0.05$ GHz (white circle), with respective $|q|$ values of 5, 1, and 0.2.}\label{fig4}
\end{figure}

\subsubsection{Effect of coupling strength on Fano features}
Furthermore, it is necessary to iteratively restructure the device to design distinct $t_{L}$ with their corresponding $L$ of the IDTs. This is crucial to manipulate the Fano resonance profile, which is not solely adjustable through $t_{L}$, but also through the variation of the coupling strength to influence $q$. Initially, if we fix the $\gamma_{\text{m}}$ and modify the $\gamma_{\text{s}}$, the Fano profiles show marginal variations due to the negligible change in $q$, as displayed in Fig.\ref{fig4}~(a). Interestingly, as we escalate $\gamma_{\text{m}}$ from weak to strong coupling [specifically, from $\gamma_{\text{m}}/2\pi=10^{-3}$ to $5\times10^{-2}$ (GHz) with a fixed $\gamma_{\text{m}}/2\pi=10^{-2}$ (GHz)], we observe the transformations in the Fano profiles. They transit from the periodic Lorentz-like shapes to the asymmetric Fano-like shapes, ultimately evolving into periodic dips. These transformations correspond to the Fano parameter moving from $|q|=5, |q|=1$ to $|q|\approx0.2$ as depicted in Fig.\ref{fig4}(b), (c), and (d), respectively. Notably, by employing $\partial_{\gamma_{\text{m}}}[q]=0$, we can derive the analytical prerequisite of $\gamma_{\text{m}}^{\text{Lor}}$ for attaining $q_{\max}$. This gives rise to the manifestation of a periodic Lorentz-like shape in the SAW scattering spectra, which can be expressed as
\begin{equation}
\begin{aligned}
\gamma _{\text{m}}^{\text{Lor}}=\frac{\sqrt{f_{\text{s1}}^4-3 t_{L}^2 \Delta
   _{\text{ep}}^2 f_{\text{s0}}^2}-f_{\text{s1}}^2}{3 T f_{\text{s0}}}.
\end{aligned}
\end{equation}
Here, for the sake of simplicity, we define $f_{\text{s0}}=\gamma_{\text{s}}t_{L}$ and $f_{\text{s1}}=f_{\text{s0}}+1$. Importantly, we can analytically determine the precise value of $\gamma_{\text{m}}^{\text{Fan}}$ that is essential for the manifestation of a periodic Fano profile within the SAW scattering spectra. This determination is accomplished by resolving Eq.~(\ref{q1}),
\begin{equation}
\begin{aligned}
\gamma _{\text{s}}^{\text{Fan}}=\frac{1}{6 T f_{\text{s0}}}\Big\{
\frac{1+\sqrt{3}i}{h_{\text{s}}}
\left[f_{\text{s1}}^4-3 \left(T\Delta _{\text{ep}}f_{\text{s0}}\right)^2\right]
+\left(1-\sqrt{3}i\right) h_{\text{s}}-2 f_{\text{s1}}^2\Big\},
\end{aligned}
\end{equation}
where $h_{\text{s}}=(3 \sqrt{f_{\text{s}}}+g_{\text{s}}+1)^{1/3}$ with the defined parameters $f_{\text{s}}=3(T \Delta_{\text{ep}} f_{\text{s0}})^6
+33 (T \Delta _{\text{ep}} f_{\text{s0}} f_{\text{s1}})^4
-3(T \Delta _{\text{ep}} f_{\text{s0}}f_{\text{s1}}^4)^2$ and $g_{\text{s}}=f_{\text{s0}}^6+6 f_{\text{s0}}^5
+3(5-6 t_{L}^2 \Delta_{\text{ep}}^2)(f_{\text{s0}}^4+f_{\text{s0}}^2) 
+4(5-9 t_{L}^2 \Delta_{\text{ep}}^2)f_{\text{s0}}^3 +6 f_{\text{s0}}$, for the sake of simplification. Nevertheless, as the value of $\gamma_{\text{m}}$ continues to escalate, and given that $q\propto 1/\sqrt{\gamma_{\text{m}}^{3}t_{L}^{2}}$, the parameter $q$ diminishes, yielding periodic dips within the SAW scattering spectrum. This enables us to modulate the periodic Fano characteristics of SAWs through strategic manipulation of the interaction between a superconducting qubit and the microwave waveguide.

\section{Summary and conclusions}
In summary, we present an innovative system for regulating periodic Fano resonances within quantum acoustic waves. This control is achieved through the strategic connection of a transducer to a ‘‘giant atom", itself coupled to a microwave waveguide. As an incident microwave photon traverses the waveguide, it incites transitions within the superconducting qubit. These transitions prompt the giant atom to produce a SAW phonon, which scatters via an IDT. This scattering process reveals distinct periodic Fano resonance characteristics within the SAW's spectrum.

To gain a deeper analytical insight into these Fano resonance features, we employ the Taylor expansion approach. This method allows us to extract the Fano parameter, a key determinant in characterizing the form and properties of Fano resonances and signifying the interference occurring between the direct and indirect scattering pathways.

Our analytical exploration uncovers that these periodic Fano resonances emerge due to the additive interference effects invoked by the dual coupling points existing between the giant atom and the acoustic wave waveguide. This understanding grants us the capability to manipulate the Fano parameter by adjusting the spacing ($L$) between these paired coupling points, which effectively changes the intrinsic time delay. By controlling the Fano parameter, we can govern the Fano profile, thereby inducing transitions between Fano and quasi-Lorentzian scattering properties, particularly in the vicinity of the resonance frequency of the giant atom's transition. Additionally, we have also established a robust theoretical framework for accurately modeling and directing the spectral width of the Fano resonances by fine-tuning the intrinsic time delay.

Furthermore, our research exemplifies that the careful adjustment of the coupling strength with the microwave waveguide, in line with our analytical model, facilitates the modulation of the entire periodic Fano resonance features from Lorentz, Fano to quasi-Lorentz shapes, not just the near-resonant Fano resonances. This novel insight allows us to precisely control the scattering behaviors and transitions between distinct resonance profiles. Our study elucidates a potent avenue for the regulation and manipulation of Fano resonances in quantum acoustic wave systems. The ramifications of our findings could have far-reaching impacts, offering exciting prospects in the realm of quantum information processing, sensing, and communication.

\begin{backmatter}
\bmsection{Acknowledgments}
YNC acknowledges the support of the National Science and Technology Council, Taiwan (MOST Grants No. 111-2123-M-006-001). This work is supported partially by the National Center for Theoretical Sciences.
\end{backmatter}

\bibliography{sample}






\end{document}